\documentclass[12pt]{iopart}
\usepackage{graphicx}

\begin{document}

\title[Kramers-Kronig relations for plasma-like  permittivities]
{Kramers-Kronig relations for plasma-like  
permittivities and the Casimir force
}
\author{
G~L~Klimchitskaya$^{1}$,
 U~Mohideen$^{2}$
and V~M~Mostepanenko$^{3}$}

\address{$^1$
North-West Technical University, Millionnaya St. 5,
St.Petersburg, Russia
}
\address{$^2$
Department of Physics and Astronomy, University of California, Riverside,
CA 92521, USA}
\address{$^3$
Noncommercial Partnership  ``Scientific Instruments'', 
Tverskaya St. 11, Moscow, Russia.}

\begin{abstract}
The Kramers-Kronig relations are derived for the permittivity
of the usual plasma model which neglects dissipation and of a generalized 
 model which takes into account the interband transitions.
The generalized plasma model is shown to be consistent with all
precision experiments on the measurement of the Casimir force.
\end{abstract}

\section{Introduction}

It is common knowledge that the Kramers-Kronig relations connect
real and imaginary parts of an analytic function describing some
causal physical process. In statistical physics and
electrodynamics any material susceptibility satisfies
the Kramers-Kronig relations \cite{1,2}.
Specifically, they are used to calculate the real part of the
dielectric permittivity, $\varepsilon^{\prime}(\omega)$,
along the real frequency axis and the dielectric permittivity,
$\varepsilon(i\xi)$, along the imaginary frequency axis \cite{2,2a}.
Both are expressed through the imaginary part of the
permittivity, $\varepsilon^{\prime\prime}(\omega)$,
at all real frequencies.

In the last few years the Kramers-Kronig relations have been
repeatedly used to calculate the thermal Casimir force
in the framework of the Lifshitz
theory (see, e.g., reviews \cite{3,4}
and recent proceedings \cite{5}). The Casimir force \cite{6}
acts between neutral material bodies and originates from the
zero-point oscillations of the electromagnetic field. 
The Lifshitz theory allows one to express the Casimir force
at a temperature $T$ in terms of $\varepsilon(i\xi)$ of the
body materials at discrete Matsubara frequencies
$\xi_l=2\pi k_BTl/\hbar$, where $k_B$ is the Boltzmann
constant and $l=0,\,1,\,2,\,\ldots\,$.
The tabulated optical data for the complex index of
refraction and hence for $\varepsilon^{\prime\prime}(\omega)$
are, however, available only in some restricted frequency
regions \cite{7}. Therefore the direct calculation of
$\varepsilon(i\xi)$ using the Kramers-Kronig relations 
is not possible and different approaches to
find $\varepsilon(i\xi)$ have been proposed.

In the first approach \cite{8,8a} the quantity
$\varepsilon^{\prime\prime}(\omega)$ obtained from the 
tabulated optical data is extrapolated using the imaginary part of
the Drude dielectric function to all lower frequencies
including zero frequency. Then $\varepsilon(i\xi_l)$ at
all $l\geq 0$ is found from the Kramers-Kronig relations and
is substituted into the Lifshitz formula at nonzero temperature. 
From the theoretical point of view this approach may seem
straightforward, but it leads to a
violation of the Nernst heat theorem for perfect crystal
lattices with no impurities \cite{9} and to contradiction
with the experiment measuring the Casimir force at
separations from 160 to 750\,nm \cite{10}. The second approach
\cite{10,11} is based on the concept of the Leontovich surface impedance
\cite{2}. This approach leads to practically the same
contributions to the Casimir force, as the first approach,
at all Matsubara frequencies with $l\geq 1$. The contribution
of the zero-frequency term is, however, different and fixed by
the impedance used. As a consequence the second approach is
not applicable at separations below the plasma wavelength (equal 
to 137\,nm for Au) where the Leontovich impedance boundary
condition becomes invalid. The third approach \cite{12,13}
does not use the tabulated optical data but employs 
the dielectric permittivity of the free electron
plasma model at all frequencies. 
Both the second and the third approaches are
in agreement with thermodynamics. They are also consistent
with the experiment \cite{10} performed at separations above
the plasma wavelength. However, both the second and third 
approaches cannot be applied in the
experiment \cite{13a,13b} where measurements start at
short separations of 60\,nm. 
This experiment although performed at $T=300\,$K 
was found to be consistent with the Lifshitz theory at
zero temperature. (The comparative analysis of all approaches is
contained in [18--20].) Note that the third, plasma 
model approach, may seem to be in disagreement with the
Kramers-Kronig relations because the dielectric permittivity
of the plasma model is entirely real. In this connection the plasma
model approach has been criticized \cite{16} for the complete
neglect of dissipation. Thus at the moment none
of the theoretical approaches to the thermal Casimir force is
consistent with all the available experimental information.

In the present paper we derive the generalized Kramers-Kronig
relations for the permittivities of 
the free electron plasma and a
plasma-like model which incorporates dissipation due to interband
transitions. We demonstrate that the permittivity of
the plasma model (as any function analytic in the upper
half-plane) satisfies the Kramers-Kronig relations if the
contribution from the pole of the second order at zero frequency is
correctly taken into account. Then we compare theoretical
computations of the thermal Casimir force using the 
free-electron plasma model and the generalized plasma model
incorporating interband transitions with the zero-temperature
Casimir force calculated using the tabulated optical data.
We demonstrate that the theoretical results using the
generalized plasma model are in good agreement with
experiment. Thus currently it is the only model for the
thermal Casimir force which is consistent with all measurements 
performed to date. We conclude with a discussion of different 
types of dissipation processes and their role in the theoretical 
description of the Casimir force.

\section{Kramers-Kronig relations for plasma and plasma-like models}

We consider the generalized plasma-like dielectric permittivity
of the form \cite{18}
\begin{equation}
\varepsilon(\omega)-1=A(\omega)-\frac{\omega_p^2}{\omega^2},
\label{eq1}
\end{equation}
\noindent
where $\omega_p$ is the plasma frequency and the oscillator term
\begin{equation}
A(\omega)=\sum\limits_{j=1}^{K}
\frac{f_j}{\omega_j^2-\omega^2-i g_j\omega}
\label{eq2}
\end{equation}
\noindent
takes into account the interband transitions of core electrons. 
Here $\omega_j\neq 0$ are the resonant frequencies of the core
electrons, $g_j$ are the respective relaxation
frequencies, $f_j$ are the oscillator strengths,
and $K$ is the number of oscillators.
The dielectric permittivity (\ref{eq1}), (\ref{eq2}) was used
in Sec.~7.5(D) of \cite{18} for the description of a metal at
frequencies much larger than the Drude relaxation frequency.
The term $-\omega_p^2/\omega^2$ in Eq.~(\ref{eq1}) describes the 
free conduction electrons and leads to a
purely imaginary current. This contribution to
$\varepsilon(\omega)$ is entirely real and does 
not include dissipation. It must be emphasized that the
oscillator term (\ref{eq2}) does not include the oscillator with
zero resonant frequency $\omega_0=0$. Thus it does not
describe conduction electrons but only the core electrons.
If the core electrons were excluded from our consideration then
$f_j=0$, $A(\omega)=0$,  and the dielectric
permittivity (\ref{eq1}) leads to the usual plasma model.
Note that for the purpose of the computations below we follow
 the notations from \cite{18a} (Level 2,\,D) for the parameters 
of the interband oscillators. Because of this we have replaced the
relaxation parameter $\Gamma_j$ in \cite{18} for $g_j$ and the
oscillator strengths $4\pi Ne^2f_j/m$, where $N$ is the number
of molecules per unit volume, as in \cite{18}, for $f_j$.
Here we also use $1+A(\omega)$ in place of
$\varepsilon_b(\omega)$ \cite{18}.
Equations (\ref{eq1}) and (\ref{eq2}) incorporate dissipation
due to interband transitions but do not include processes
of electron scattering on phonons, impurities, grain
boundaries, surfaces and other electrons. 
Below we investigate the mathematical properties of Eqs.~(\ref{eq1}), 
(\ref{eq2}) for the complete frequency range from zero to infinity. 
The physical justification for the choice of
$\varepsilon(\omega)$ in Eq.~(\ref{eq1}) is discussed in Sec.~4.

\begin{figure*}[b]
\vspace*{-13.4cm}
\includegraphics{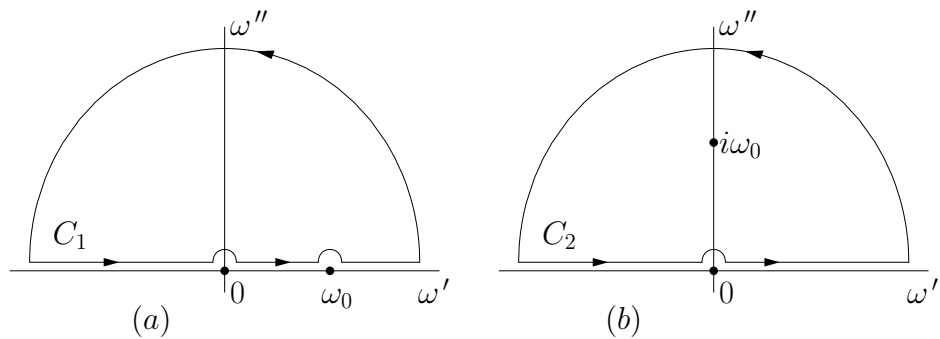}
\vspace*{-12.6cm}
\caption{
The integration contours ($a$) in Eq.~(\ref{eq3}) and ($b$) 
in Eq.~(\ref{eq11}) consisting of the real frequency axis and
the semicircle of infinitely large radius.
}
\end{figure*}
The characteristic feature of the dielectric permittivity (\ref{eq1})
is the second order pole at zero frequency. Let us demonstrate
that (\ref{eq1}) satisfies the Kramers-Kronig relations in both
cases $A(\omega)=0$ and $A(\omega)\neq 0$. For this purpose we
consider the integral
\begin{equation}
\int_{C_1}\frac{\varepsilon(\omega)-1}{\omega-\omega_0}
d\omega=0,
\label{eq3}
\end{equation}
\noindent
where $\omega_0$ is real and the contour $C_1$ is presented in Fig.~1($a$).
Inside $C_1$ the function under the integral is analytic and thus the
equality (\ref{eq3}) follows from the Cauchy theorem. At infinity
$\varepsilon(\omega)\to 1$ and the function
$[\varepsilon(\omega)-1]/(\omega-\omega_0)$ therefore tends to
zero more rapidly than $1/\omega$. Because of this the integral
along the semicircle of infinite
radius is zero. We pass around the
points 0 and $\omega_0$ along the semicircles $C_{\rho}$ and $C_{\delta}$ 
with radii $\rho$ and $\delta$, respectively. It is easily seen that
\begin{equation}
\int_{C_{\delta}}\frac{\varepsilon(\omega)-1}{\omega-\omega_0}
d\omega=-\pi i\mbox{Res}\frac{\varepsilon(\omega)-1}{\omega-\omega_0}
\vert_{\omega=\omega_0}=-\pi i\left[\varepsilon(\omega_0)-1\right].
\label{eq4}
\end{equation}
\noindent
The similar integral around the point 0 is more involved.
Using Eq.~(\ref{eq1}) we represent it as a sum of the integral
\begin{equation}
\int_{C_{\rho}}\frac{A(\omega)}{\omega-\omega_0}
d\omega=-\frac{2A(0)}{\omega_0}\rho,
\label{eq5}
\end{equation}
\noindent
which vanishes when $\rho\to 0$, and
\begin{equation}
-\omega_p^2\int_{C_{\rho}}\frac{d\omega}{\omega^2(\omega-\omega_0)}
\equiv\frac{\omega_p^2}{\omega_0^2}\int_{C_{\rho}}
\left[\frac{\omega_0}{\omega^2}-\frac{1}{\omega-\omega_0}+
\frac{1}{\omega}\right]d\omega.
\label{eq6}
\end{equation}
\noindent
Direct integration along the semicircle $C_{\rho}$ results in
\begin{eqnarray}
&&
\int_{C_{\rho}}\frac{d\omega}{\omega-\omega_0}=
-\frac{2}{\omega_0}\rho, \qquad
\int_{C_{\rho}}\frac{d\omega}{\omega}=-\pi i,
\nonumber \\
&&
\int_{C_{\rho}}
\frac{d\omega}{\omega^2}=-\frac{2}{\rho}=\omega_0
P\!\int_{-\infty}^{\infty}
\frac{d\omega}{\omega^2(\omega-\omega_0)},
\label{eq7}
\end{eqnarray}
\noindent
where the integral is taken as a principal value.
[Note that the last integral cannot be evaluated as in
Eq.~(\ref{eq4}) because Res$(1/\omega^2)\vert_{\omega=0}=0$ and
both integrals around the upper and lower semicircles are
divergent and opposite in sign.]

Substituting Eqs.~(\ref{eq4})--(\ref{eq7}) in Eq.~(\ref{eq3})
we arrive at
\begin{equation}
-\frac{i\pi\omega_p^2}{\omega_0^2}-i\pi\left[\varepsilon(\omega_0)
-1\right]+P\!\int_{-\infty}^{\infty}
\frac{d\omega}{\omega-\omega_0}\left[\varepsilon(\omega) -1
+\frac{\omega_p^2}{\omega^2}\right]=0.
\label{eq8}
\end{equation}
\noindent
Now we replace the integration variable $\omega$ by $\xi$,
$\omega_0$ by $\omega$, and represent the function
$\varepsilon(\omega)$ in the form of
$\varepsilon(\omega)=\varepsilon^{\prime}(\omega)+
i\varepsilon^{\prime\prime}(\omega)$. Taking into account that
\begin{equation}
P\!\int_{-\infty}^{\infty}
\frac{d\omega}{\omega-\omega_0}=0
\label{eq9}
\end{equation}
\noindent
and separating the real and imaginary parts in Eq.~(\ref{eq8}),
we obtain the generalized Kramers-Kronig relations
\begin{equation}
\varepsilon^{\prime}(\omega)=1+\frac{1}{\pi}
P\!\int_{-\infty}^{\infty}
\frac{\varepsilon^{\prime\prime}(\xi)}{\xi-\omega}d\xi-
\frac{\omega_p^2}{\omega^2},
\quad
\varepsilon^{\prime\prime}(\omega)=-\frac{1}{\pi}
P\!\int_{-\infty}^{\infty}
\frac{\varepsilon^{\prime}(\xi)+\frac{\omega_p^2}{\xi^2}}{\xi-\omega}d\xi.
\label{eq10}
\end{equation}
Note that 
the standard relations \cite{2} obtained for permittivities with no
pole at $\omega=0$ do not contain terms $\omega_p^2/\xi^2$ on the
right-hand sides of Eqs.~(\ref{eq10}). 

The dielectric permittivity along the imaginary frequency axis can be 
determined through the use of the integral
\begin{equation}
\int_{C_2}
\frac{\omega\left[\varepsilon(\omega)-1\right]}{\omega^2+\omega_0^2}
d\omega=\pi i\left[\varepsilon(i\omega_0)-1\right] 
\label{eq11}
\end{equation}
\noindent
along the contour $C_2$ in Fig.~1($b$). By integrating over $C_2$
we get 
\begin{equation}
i\pi\frac{\omega_p^2}{\omega_0^2}+
P\!\int_{-\infty}^{\infty}
\frac{\omega\left[\varepsilon(\omega)-1\right]}{\omega^2+\omega_0^2}
d\omega=i\pi\left[\varepsilon(i\omega_0)-1\right]. 
\label{eq12}
\end{equation}
\noindent
Now we make the same replacement of variables as above, separate the real 
and imaginary parts of $\varepsilon(\omega)$ under the integral and use
the identities
\begin{equation}
P\!\int_{-\infty}^{\infty}
\frac{\xi d\xi}{\xi^2+\omega^2}=0, \qquad
P\!\int_{-\infty}^{\infty}
\frac{\xi \varepsilon^{\prime}(\xi)}{\xi^2+\omega^2}d\xi=0.
\label{eq13}
\end{equation}
\noindent
The result is
\begin{equation}
\varepsilon(i\omega)-1=\frac{1}{\pi}
P\!\int_{-\infty}^{\infty}
\frac{\xi \varepsilon^{\prime\prime}(\xi)}{\xi^2+\omega^2}d\xi
+\frac{\omega_p^2}{\omega^2}.
\label{eq14}
\end{equation}

For the usual plasma model  
$\varepsilon^{\prime\prime}(\omega)=0$,
$\varepsilon^{\prime}(\omega)=1-\omega_p^2/\omega^2$,
$\varepsilon(i\omega)=1+\omega_p^2/\omega^2$,
the generalized Kramers-Kronig relations (\ref{eq10}),
(\ref{eq14}) are  
satisfied with the use of Eq.~(\ref{eq9}).
On the contrary the same plasma model violates the standard
Kramers-Kronig relations.
Note that sometimes \cite{19} the plasma model is ascribed
a nonzero imaginary part
\begin{equation}
\varepsilon^{\prime\prime}(\omega)=-\frac{\omega_p^2}{\omega}
\lim\limits_{g\to 0}\frac{1}{\omega+ig}=
\frac{\omega_p^2}{\omega}\pi\delta(\omega),
\label{eq15}
\end{equation}
\noindent
which is obtained from the Drude dielectric function in the
limit of zero relaxation parameter. This makes it possible
to formally satisfy the standard dispersion relation for
the dielectric permittivity along the imaginary frequency
axis \cite{2} given by Eq.~(\ref{eq14}) without the
$\omega_p^2/\omega^2$ term.
However the other two standard Kramers-Kronig relations with
$\varepsilon^{\prime\prime}(\omega)$, as given by Eq.~(\ref{eq15}),
become meaningless. Thus the permittivity of the collisionless free
electron gas is entirely real.

It is easily seen that the plasma-like dielectric permittivity (\ref{eq1}),
(\ref{eq2}) satisfies the generalized Kramers-Kronig relations 
(\ref{eq10}) and (\ref{eq14}). This can be verified by direct
substitution. For example, the substitution of Eqs.~(\ref{eq1}) and
(\ref{eq2}) in the first equation of 
(\ref{eq10}) leads to
\begin{eqnarray}
&&
\sum\limits_{j=1}^{K}
\frac{f_j\left(\omega_j^2-\omega^2\right)}{\left(\omega_j^2-
\omega^2\right)^2+g_j^2\omega^2}=\frac{1}{\pi}
\sum\limits_{j=1}^{K}f_jg_j
P\!\int_{-\infty}^{\infty}
\frac{\xi d\xi}{(\xi-\omega)\left[\left(\omega_j^2-\xi^2\right)^2+
g_j^2\xi^2\right]}
\nonumber \\
&&\phantom{aa}
=\frac{1}{\pi}
\sum\limits_{j=1}^{K}\frac{f_jg_j}{\left(\omega_j^2-\omega^2\right)^2+
g_j^2\omega^2}
\label{eq17} \\
&&\phantom{aaaaaa}\times
\left[\omega_j\int_{-\infty}^{\infty}
\frac{dy}{y^4-2\beta_jy^2+1}-
\frac{\omega^2}{\omega_j}\int_{-\infty}^{\infty}
\frac{y^2dy}{y^4-2\beta_jy^2+1}\right],
\nonumber
\end{eqnarray}
\noindent
where $\beta_j\equiv 1-g_j^2/(2\omega_j^2)$.
When the following:
\begin{equation}
\int_{-\infty}^{\infty}\frac{dy}{y^4-2\beta_jy^2+1}=
\int_{-\infty}^{\infty}\frac{y^2dy}{y^4-2\beta_jy^2+1}=
\frac{\pi}{\sqrt{2(1-\beta_j)}}
\label{eq18}
\end{equation}
\noindent
is taken into account
Eq.~(\ref{eq17}) is satisfied. The second equation
in (\ref{eq10}) and Eq.~(\ref{eq14}) can be verified in a similar way.

\section{Calculation of the Casimir force using the 
generalized plasma model and comparison with experiment}

We have calculated the thermal Casimir force $F^{\rm th}$ acting
between Au coated test bodies in the most precise
short-separation experiment \cite{13a}
(a sphere of $R=95.65\,\mu$m radius and a plate) by substituting
the generalized plasma dielectric function (\ref{eq1}), (\ref{eq2})
into the Lifshitz formula. The Lifshitz formula for the force
between a sphere and a plate was obtained by means of the
proximity force approximation (PFA) as $2\pi R{\cal F}$ where ${\cal F}$ 
is the free energy in the configuration of two parallel plates
(see \cite{4,13b} for details). 
Note that recently the high accuracy of PFA was confirmed using the
path-integral approach \cite{pfa1,pfa2} and worldline numerics \cite{pfa3}.
The oscillator parameters for Au
in Eq.~(\ref{eq2}) were found in \cite{18a,20} using
DESY data. They are as follows: $K=3$, 
$\omega_1=3.87\,$eV, $f_1=59.61\,(\mbox{eV})^2$, $g_1=2.62\,$eV,
$\omega_2=8.37\,$eV, $f_2=122.55\,(\mbox{eV})^2$, $g_2=6.41\,$eV,
$\omega_3=23.46\,$eV, $f_3=1031.19\,(\mbox{eV})^2$, $g_3=27.57\,$eV.
Computations were performed at the laboratory temperature
$T=300\,$K at different experimental separations $a$ with
$\omega_p=9.0\,$eV \cite{7,13b}. The obtained magnitudes of the
Casimir force are presented in Table~1 (column 2).
\begin{table}[b]
\caption{
Magnitudes of the Casimir force at different separations in column 1
computed at $T=300\,$K using the generalized plasma-like model
(column 2), the usual plasma model (column 3), and at zero
temperature using the tabulated optical data for the complex
index of refraction (column 4).
}
\begin{indented}
\item[]
\begin{tabular}{@{}cccc}
\br
&\centre{3}{Force magnitude (pN)}\\
\ns&\crule{3}\\
$a$ & Generalized & Plasma & Force at zero \\
(nm) & plasma model& model& temperature \\
\mr
60 & 531.1 & 483.2 & 527.4 \\
70 & 358.8 & 332.2 & 356.1 \\
80 & 254.9 & 239.1 & 252.8 \\
90 & 188.2 & 178.3 & 186.5 \\
100 & 143.3 & 136.8 & 141.9 \\
120 & 88.94 & 86.00 & 88.01 \\
150 & 49.30 & 48.19 & 48.71 \\
200 & 22.75 & 22.46 & 22.44 \\
250 & 12.37 & 12.28 & 12.19 \\
300 & 7.478 & 7.438 &7.355 \\
\br
\end{tabular}
\end{indented}
\end{table}
For comparison in column 3 we present the force magnitudes 
obtained using the usual plasma model of Eq.~(\ref{eq1}) with 
$A(\omega)=0$. Column 4 lists the force magnitudes 
from the zero-temperature Lifshitz formula  using the
tabulated optical data for the complex index of refraction (recall
that in \cite{13a,13b} the experimental data were compared with
theory at zero temperature). As is seen from Table~1, at short
separations the results from the usual plasma
model (column 3) deviate significantly from the predictions of
the generalized plasma model (column 2). At the same time
the results in column 2 are in close agreement with
computations at $T=0$ (column 4) which are both consistent with
experiment.

\begin{figure*}[t]
\vspace*{-14cm}
\includegraphics{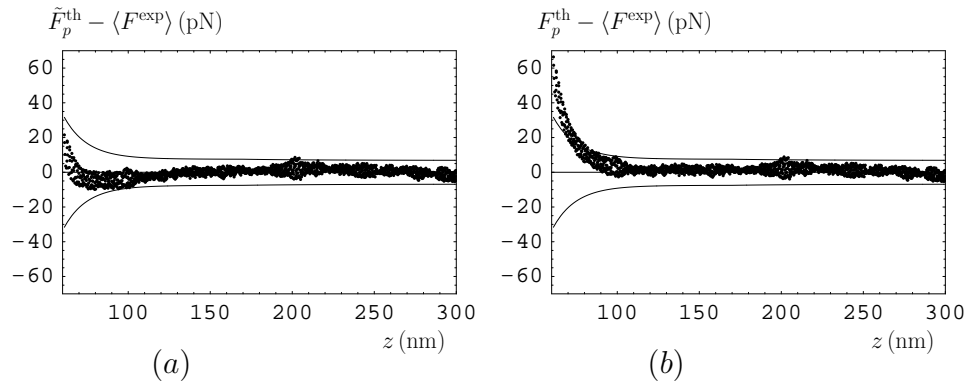}
\vspace*{-11.3cm}
\caption{
Differences between the theoretical and mean experimental \cite{13a}
Casimir forces versus separation. Theoretical forces are
computed using ($a$) the generalized plasma-like model and ($b$) the
usual plasma model.
}
\end{figure*}
Direct comparison with the experimental data of \cite{13a}
confirms that the generalized plasma model (\ref{eq1}), (\ref{eq2})
is consistent with measurements of the Casimir force at short separations.
In Fig.~2($a$) we plot the differences between the theoretical forces
computed using the generalized plasma model, ${\tilde{F}}_p^{\rm th}$,
and the mean values of the measured forces. As is seen in Fig.~2($a$),
almost all points are inside the error bars for the
difference between theory and experiment \cite{10,21} computed at a
95\% confidence level (solid lines). Notice that the comparison of
data with the zero-temperature theoretical force practically
coincides with that shown in Fig.~2($a$). By contrast in Fig.~2($b$) the
theoretical force, $F_p^{\rm th}$, is computed using the usual
plasma model. It is clearly seen that at short separations the usual 
plasma model is inconsistent with data whereas at larger
separations both models are in agreement with measurements.

Refering back to the Introduction, the usual plasma model is
consistent with the data of experiment \cite{10} which is the most
precise experiment at
separations from 160 to 750\,nm. At such large separations the 
predictions of the generalized and usual plasma models almost 
coincide (in fact, the use of the generalized plasma model 
instead of the usual one leads
to slightly better agreement between experiment and theory
in \cite{10}). Thus we can conclude that the generalized 
plasma model (\ref{eq1}), (\ref{eq2}) both exactly satisfies the
Kramers-Kronig relations and is also the only model consistent
with all precise experiments on the Casimir force performed
to date.

\section{Discussion}

As was mentioned in the Introduction, the usual plasma model neglects
the role of free electron scattering. 
In the absence of scattering the conductivity of free electrons
is purely imaginary and the dielectric permittivity is completely real
signifying the absence of dissipation. The generalized plasma-like
dielectric permittivity (\ref{eq1}), (\ref{eq2}), as well as
the usual plasma model, does not take into account the scattering
processes of the free electrons. However, it includes dissipation
due to the interband transitions of core electrons. These
transitions are described by a set of oscillators with nonzero
resonant frequencies. Our results demonstrate that the
generalized plasma model exactly satisfies the Kramers-Kronig
relations and is also consistent with all available experimental
data. In the same way, as in \cite{9,22,23}, it can also be shown
that this model is in agreement with the Nernst heat theorem.

The dissipation processes of free electrons on phonons, 
impurities etc. are not included in the generalized plasma model.
 They can be taken into account by modifying
Eq.~(\ref{eq1}) to the form
\begin{equation}\varepsilon(\omega)-1=A(\omega)-
\frac{f_0}{\omega(\omega +ig_0)},
\label{eq19}
\end{equation}
\noindent
where $f_0=\omega_p^2$. The second term on the right-hand side
of Eq.~(\ref{eq19}) is the contribution of an oscillator
with zero resonant frequency, $\omega_0=0$, 
which was not included in Eq.~(\ref{eq2}).
This additional oscillator results in a first order pole
and leads to the Drude-type term in the dielectric permittivity. 
The Kramers-Kronig relations for the dielectric permittivity 
(\ref{eq19}) are familiar \cite{2}. However,
as was discussed in the Introduction, the use of the dielectric
permittivity (\ref{eq19}) leads to a violation
of the Nernst heat theorem for perfect crystal lattices and to a
contradiction with experiment \cite{10}. The question why the
inclusion of one type of dissipation (interband transitions of core
electrons) in the Lifshitz theory is necessary while that of another  
(scattering processes of free electrons) leads to contradictions with 
fundamental physical principles and experiment remains open. Future 
theoretical and experimental developments will shed light on this issue.

\section*{Acknowledgments}
This work was supported by the DOE Grant No.~DE-FG02-04ER46131. 
Numerical computations were supported by the NSF Grant
No.~PHY0355092.

\end{document}